\documentstyle[12pt]{article}

\makeatletter
\@addtoreset{equation}{section}

\makeatother

\def\beq{\begin{equation}}
\def\eeq{\end{equation}}
\def\bec{\begin{center}}
\def\eec{\end{center}}
\def\bea{\begin{eqnarray}}
\def\nn{\nonumber \\ }
\def\eea{\end{eqnarray}}

\def\ds{\displaystyle}

\def\ni{\noindent}

\def\req#1{(\ref{#1})}
\def\ie{i.e.\, }

\begin{document}

\begin{center}
{\Large\bf Renormalization Group Properties of
Higher-Derivative Quantum Gravity with Matter
in $4-\varepsilon$ Dimensions} \\

\vspace{1cm}
Emili {\sc Elizalde$^{a,b,}$}\footnote{E-mail:
eli@zeta.ecm.ub.es},
Sergei D. {\sc Odintsov$^{b,}$}\footnote{On
 leave of absence from Tomsk Pedagogical Institute,
634041 Tomsk, Russia. E-mail: odintsov@ecm.ub.es},
and August {\sc Romeo$^{a,}$}\footnote{E-mail:
august@ceab.es, august@zeta.ecm.ub.es},
\vspace{3mm}

$^a$Centre for Advanced Studies CEAB, CSIC,
Cam\'{\i} de Santa B\`arbara, 17300 Blanes \\
$^b$Department ECM and IFAE,
Faculty of Physics, University of  Barcelona, \\
Diagonal 647, 08028 Barcelona,
Catalonia, Spain \\

\vspace{15mm}

{\bf Abstract}
\eec

We investigate the phase structure and the infrared properties of
higher-derivative quantum gravity (QG)
with matter,  in $4-\varepsilon$ dimensions. The renormalization
group (RG)
equations in $4-\varepsilon$ dimensions are analysed for the following
types of matter: the $f\varphi^4$-theory, the $O(N) \varphi^4$-theory,
scalar
electrodynamics, and the $SU(2)$ model with scalars.
New fixed points for the scalar coupling appear ---one of which is IR
stable---  being some of them induced by QG.
The IR stable fixed point perturbed by QG leads to a second-order phase
transition for the theory at non-zero temperature. The RG improved
effective potential in the $SU(2)$ theory (which can be considered
as the confining phase of the standard model) is
calculated at nonzero temperature and it is shown that its shape is
clearly influenced by QG.

\newpage

\section{Introduction}

The infrared (IR) structure of quantum gravity (QG) is very important
in the applications of QG to particle physics and cosmology at scales
below the Planck mass. Regarded from another side, the IR behaviour of
QG is closely connected with its thermal properties.
For example, the study of Einstein gravity on a de Sitter background
\cite{GHP} is motivated not only by possible cosmological applications,
but also by the need to understand the IR structure of the graviton
propagator \cite{F}. IR dynamics of QG is most probably relevant to
the solution of the cosmological constant problem \cite{TV}. Further
exploiting this point of view, an effective
theory \cite{AM} based on integrated conformal anomaly dynamics \cite{DDI}
has been developed,  aimed at the description of QG in the far infrared
region.
Such a model may provide a very natural solution of the cosmological
constant problem.

To investigate the IR properties of a theory, the well-known
$\varepsilon$-expansion technique is often very useful, as it happens to
be the case in the theory
of critical phenomena (see \cite{ZJ}-\cite{BGZJ} for an introduction).
For renormalizable theories, this technique supplies a quite standard
way of studying
the critical behaviour and the nature of its phase-transitions at
nonzero
temperature \cite{Wi,G}. (A discussion on the equivalence between
 the IR
behaviour of finite-temperature theories and three-dimensional theories
can be found in \cite{AP,WR}).

In QG, the $\varepsilon$-expansion technique has been applied to study
the corresponding ($2+\varepsilon$)-dimensional theory \cite{W,JJ,NTT}.
However, standard Einstein gravity is not renormalizable, not even in
$2+\varepsilon$ dimensions \cite{JJ}. To overcome this difficulty, it
looks certainly
interesting to study a multiplicatively renormalizable QG starting from
a different standpoint in $D=4-\varepsilon$ dimensions.
Such a study
is expected to be a more realistic approach to QG than the $D=2+
\varepsilon$ theory,
since two-dimensional theories have very specific properties.

At the present moment there is no completely consistent theory
of quantum gravity. Standard Einstein gravity ---which is very fine as a
classical theory--- is regretfully non-renormalizable \cite{tHV}.
String theory (see \cite{35} for a general review) may certainly
 lead to an
effective gravity, which in general should also be considered at the
classical level (being also, as a rule, non-renormalizable). Under such
circumstances, there is still the following possibility to take into
account quantum gravity effects: to work with some effective theory of
QG which mimics some of the essential properties of the fully consistent
theory. Basically one can think of such an effective theory as of an
expansion of the (unknown) gravitational Lagrangian in powers of
curvature invariants. In principle such a theory will be, generically
speaking, non-renormalizable.  One
exceptional example is, however, four-dimensional higher-derivative
gravity, which is known to be multiplicatively renormalizable \cite{S}
(in a scheme, of course, of a fourth-order propagator). Moreover, this
theory is asymptotically free \cite{BOS}-\cite{BS}, and has
Einstein's
theory as a low-energy limit (hence, it is undistinguishable from
Einstein's theory at the classical level). Owing to the fact that the
gravitational couplings in such a theory are dimensionless, they might
show up in different situations of high-energy physics, in the standard
description of quantum field theory.

This theory of QG is certainly not free from its own problems,
what has ruled it out from the very short list of candidates for
describing QG
in a fundamental way. The main difficulty associated with any model of
higher-derivative QG is perhaps the unitarity problem at the
perturbative
level. However, a real chance of solving the unitarity problem in a
non-perturbative approach exists (see, for example, the paper by
Antoniadis and Tomboulis in Ref. \cite{JT}). One has to emphasize that
the problem of non-unitarity is of dynamical nature, so that one is
compelled to consider it with account of all kind of quantum corrections
(perturbative and non-perturbative ones). In this respect, the string
example \cite{36} ---where it happens that negative-norm states decouple
at the renormalization group (RG) fixed point--- may be a good lesson to
learn. Also, when considered as some sort of effective theory ---that
ought to be replaced with the true fundamental theory of QG at high
energies--- the loss of unitarity should not be the ultimate argument
against higher-derivative gravity.

When considered in the energy region between the GUT scale and the
Planck scale, and in interaction with some asymptotically-free GUT,
higher-derivative gravity does not, as a rule, destroy asymptotical
freedom and, hence, a totally asymptotically free theory which
represents the unification of $R^2$-gravity with a GUT \cite{BOS} can be
easily constructed. Then, one sees the appearance of one-loop quantum
gravity
corrections to scalar and Yukawa couplings, what may influence
the GUT in different respects. In the present paper ---motivated
in addition by the big activity that is going on about the study of
($2+\epsilon$)-dimensional gravity--- we suggest investigating the
influence
of the above mentioned QG corrections on the phase transitions that
occur in scalar-gauge theories, and in their infrared properties,
working with the $\epsilon$-expansion technique \cite{WK} near four
dimensions. Finally, even if one would still insist upon the point that
higher-derivative gravity is not very realistic on some ground or other
(as we
have argued before it may have some relevance to reality), the theory
to be considered here constitutes a rather nice and completely workable
example where, as we shall see, the critical dynamics can be
studied up to almost the same level as in the
absence of QG, what is already a quite appealing issue (for a list of
different theories whith similar studies of their properties in relation
with critical phenomena, see \cite{ZJ}).

The paper is organized as follows. In the next section we give the
action of QG with matter in $4-\varepsilon$ dimensions, and analyse some
of their properties. In section 3 we investigate the RG equations in the case of
higher-derivative QG interacting with  matter  in $4-\epsilon$ dimensions.
 One of the main motivations for such a
study is, of course, the search for finite-temperature phase
transitions and, hence, we would wish to set $\epsilon =1$ in the end.
Unfortunately, in general the $\epsilon$ expansion \cite{WK} works
well only near $D=4$, i.e., for infinitesimal $\epsilon$. As
usual in these cases, we will here assume  that the results rigorously
derived for small $\epsilon$ continue to be valid at $\epsilon =1$.
In other areas of physics there exist plenty of explicit
examples (see, for
instance, Ref. \cite{ZJ}) where, having at our disposal both numerical
and experimental data, the comparison of both gives unexpectedly good
agreement (especially for the case of second order phase transitions).
Regretfully, in the theory under consideration here we have yet no
numerical results to compare, for simulations of QG are very hard to do,
even in low dimensions (see Ref. \cite{34} for a recent review). But,
from the mentioned experience it would come as  no surprise if things
would really match again, when the comparison can be established.
The general structure of the solutions
and the appearance of new fixed points induced by QG (in the case of the
$f\varphi^4$-theory) are discussed, as well as their relevance to the
study of QG at nonzero temperature. Note that in earlier papers (see
\cite{GPY,BFT} and references therein) mainly Einstein QG at nonzero
temperature has been studied. In particular, the temperature corrections
to the graviton amplitudes have been discussed in Ref. \cite{BFT}.
 We will also investigate the fact that near the IR stable fixed point a second
order phase transition occurs. In the case of the $SU(2)$
model, only unstable fixed points exist (due to the fact that
$g^{* 2}=0$) and hence a first order phase transition is to be
expected.

Section 4 is dedicated to the study the RG equations in
 higher-derivative QG with the $O(N)$ model. The
phase structure is numerically investigated, showing that at large $N$
the explicit value of the fixed point decreases.
In section 5 we study the RG-improved effective potential at nonzero
temperature in a $SU(2)$ model with QG and estimate the influence
of QG on the confining phase of the standard model (SM).
We conclude with an overall discussion and an outlook.

\section{Higher-derivative QG with matter in $4-\varepsilon$ dimensions}

We will be interested in studying the behaviour of theories in
$4-\varepsilon$ dimensions (eventually $\varepsilon$ will be chosen
to be 1) with a Lagrangian of the form
\beq
{\cal L}=\mu^{-\varepsilon} \left( {1 \over 2\lambda}W
-{\omega \over 3 \lambda} R^2 + \chi R + \Lambda \right)
+{1 \over 2} \xi R \varphi^2
+ {\cal L}_m( \varphi, A^a_{\mu} ),
\label{L}
\eeq
where $W=C_{\mu \nu \alpha \beta} C^{\mu \nu \alpha \beta}$ is the
square of the Weyl tensor, $\varphi$ is a scalar and $\mu$ is a mass
parameter which is used to make $\lambda$
dimensionless in $4-\varepsilon$ dimensions.
The symbolic form for the matter Lagrangian is
\beq
{\cal L}_m( \varphi, A^a_{\mu} )=
-{1 \over 4}G^a_{\mu \nu} G^{a \mu \nu}
+{1 \over 2} g^{\mu \nu} (\nabla_{\mu}\varphi)^a (\nabla_{\nu}\varphi)^a
-{1 \over 2} m^2 \varphi^2
-{1 \over 4!}f \mu^{\varepsilon} \varphi^4,
\label{Lm}
\eeq
where $\varphi^2=\varphi^a \varphi^a$, $(\nabla_{\mu}\varphi)^a$ is a
covariant derivative in which the standard gauge coupling
$g$ has been changed in the way $g \to \mu^{\varepsilon/2}g$, in
order to keep $g$ dimensionless for the number of dimensions
$D=4-\varepsilon$. The same is true for $f$.

To study the critical behaviour of such a system, which includes a
higher-derivative QG sector plus the (non)-abelian Higgs model (for an
introduction to critical phenomena, see \cite{ZJ}-\cite{BGZJ}), it
is enough
to consider the renormalizable subset of the theory \cite{G}
where
masses are extremely small or zero. So, we may put $m^2=0$ in \req{Lm}.
We shall later discuss the different content of fields and different
gauge groups in ${\cal L}_m$ \req{Lm}.

It is worth noting that a detailed introduction to higher-derivative QG
(with matter) was given in the book \cite{BOS} (see also refs. therein).
As mentioned in the introduction, such a theory
is known to be multiplicatively renormalizable in $D=4$ and also
asymptotically free for all coupling constants. Asymptotic freedom
in the matter sector may be also realized \cite{BOS,BS} (if the gauge
couplings
show an AF behaviour, \ie if abelian vectors are not included in the
theory), hence the QG couplings play in some sense the role
of a Yukawa coupling constant.

The study of the RG behaviour of the dimensional parameters
$( \chi, \Lambda )$ ---which play the role of some kind of mass for
the gravitational field--- is of less interest to us. First of all, the
RG equations for these couplings are gauge dependent and,
in order to render them gauge-independent, one has to introduce some
combination (an essential coupling constant). The RG equation
for this essential coupling constant is certainly gauge-independent.
Moreover, without the cosmological and Einstein terms, the theory
\req{L}
is multiplicatively renormalizable if the scalar mass in \req{Lm}
is zero.
Applying the same arguments
as in \cite{Wi,G} ---namely that in order to study its critical
behaviour
it is enough to consider the effective renormalizable subset of the
theory with zero masses (which is indeed the case!)---
one can actually drop these
terms from the very beginning. So, we shall concentrate only
on the
study of the terms which have dimensionless coupling constants in four
dimensions.

It is interesting to observe that even if one starts from the theory 
without Einstein or
cosmological
terms, one may recover them in the low-energy theory as a result of
spontaneous symmetry breakdown (or a curvature-induced phase transition
\cite{BO}).
Note also that in what follows we assume a topologically trivial,
flat background for the gravitational field.

\section{Renormalization group equations for $R^2$-gravity with matter}

In this section we will consider the RG  equations using
 the $\varepsilon$-expansion analysis. As specific
examples, we take
a few different matter theories, in particular the $f\varphi^4$-theory,
scalar electrodynamics, and the nonabelian $SU(2)$ model.
Observe that we will work in frames of standard perturbation theory,
 where QG corrections are essential.

Let us consider the $f\varphi^4$-theory in flat
$D=4-\varepsilon$ dimensional spacetime. Then,
\beq
{\cal L}_m={1 \over 2}\partial^{\mu}\varphi\partial_{\mu}\varphi
-{1 \over 4!}f \mu^{\varepsilon} \varphi^4.
\label{Lmfphi4}
\eeq
As usual, in $4-\varepsilon$ dimensions the RG flows are generated
by the RG equation:
\beq
{df \over dt}=-\varepsilon f +\beta_f(f),
\label{RGdf}
\eeq
where $\beta_f$ is the four-dimensional $\beta$-function in the theory
\req{Lmfphi4} and we are interested in its $t \to -\infty$
behaviour.
Indeed, it is the IR behaviour of the theory
what dertermines the critical features of the system.
The one-loop $\beta$-function for the $f\varphi^4$-theory is
well known:
\beq \beta_f={3 f^2 \over (4 \pi)^2} \eeq
As one can see, apart from $f^*=0$ ---which is unstable--- there is a
stable fixed point at
\beq f^*={ (4 \pi)^2 \varepsilon \over 3}. \label{fs}\eeq
(of course, the value of $f^*$ is quite large).
This is a quite known result (see, for example \cite{BGZJ,ZJ})
which shows that such a model near a phase transition gives a
second-order
transition behaviour. The solution of \req{RGdf} may be explicitly
written:
\beq
f(t)={ f \varepsilon (4 \pi)^2 \over
3f-(3f- \varepsilon (4 \pi)^2) e^{\varepsilon t} }.
\eeq
Clearly, as $t \to -\infty$, $f(t) \to f^*$.
As is well known, first order phase transitions are very common in
quantum field theory. We will also find that, in the system under
discussion here, mainly first order phase trasitions occur but, in some
cases we will obtain second order ones, what will be certainly a
nice result. In fact, first order phase transitions are of lesser
physical interest, in the mean field theory language \cite{ZJ}: the
correlation length corresponding to such systems is finite and no
universality emerges. As it was mentioned long ago \cite{G}, if the
theory possesses no stable fixed point in the $\epsilon$ expansion then
the first order phase transitions that are predicted are, in general,
only weakly of first order (some other effects can become important, as
well). Concerning the more physically interesting second order phase
transitions, they are not so common, as mentioned already. In the
$\epsilon$-expansion technique we are mainly interested in $\epsilon
=1$, i.e. in finite temperature systems. And it was pointed out in
\cite{ZJ}-\cite{BGZJ} that for second order phase transitions the lowest
order results at $\epsilon$,    often   agree very well with the
experiments (even at $\epsilon =1$), when they are available.

The first example will be  scalar self-interacting theory 
with higher derivative gravity. In this case the
system of RG equations is
\beq
\begin{array}{lllll}
\ds{d\lambda\over dt}&=&\beta_{\lambda}&=&\ds-\varepsilon\lambda
-{799 \lambda^2 \over 60 (4 \pi)^2}, \\
\ds{d\omega\over dt}&=&\beta_{\omega}&=&\ds
-{\lambda \over (4 \pi)^2}\left[ {10 \over 3}\omega^2
+\left( 18+{19 \over 60} \right) \omega + {5 \over 12}
+{3 \over 2} \left( \xi -{1 \over 6} \right)^2 \right], \\
\ds{d\xi\over dt}&=&\beta_{\xi}&=&\ds {1 \over (4 \pi)^2}
\left[ f\left( \xi -{1 \over 6} \right)
+\lambda\xi\left(
-{3 \over 2}\xi^2 +4\xi +3 +{10 \over 3}\omega
-{9 \over 4 \omega}\xi^2 +{5 \over 2\omega}\xi- {1 \over 3\omega}
\right) \right] \\
&&&\equiv&\beta_{\xi}^{(0)}+\Delta\beta_{\xi}, \\
\ds{df\over dt}&=&\beta_f&=&\ds-\varepsilon f
+{1 \over (4 \pi)^2}
\left[ 3f^2 +\lambda^2\xi^2\left( 15+ {3 \over 4\omega^2}
-{9 \xi \over \omega^2}+{27 \xi^2 \over \omega^2} \right) \right. \\
&&&&\hspace*{6em}\ds \left. -\lambda f\left(
5+3\xi^2+{33 \over 2\omega}\xi^2
-{6 \over \omega}\xi +{1 \over 2 \omega} \right) \right] \\
&&&\equiv&-\varepsilon f+\beta_f^{(0)}+\Delta\beta_f,
\end{array}
\label{betasloxf}
\eeq
where $\beta_f^{(0)}={3f^2 \over (4 \pi)^2}$,
$\beta_{\xi}^{(0)}={ f\left( \xi -{1 \over 6} \right) \over (4 \pi)^2}$,
the remaining parts of $\beta_{\xi}$ and $\beta_f$ defining
$\Delta\beta_{\xi}$ and $\Delta\beta_f$,
which are universal QG corrections to $\beta_{\xi}$ and $\beta_f$.

For $\lambda >0$ (this is the situation that is usually considered,
and then the classical action is positively defined) we have
asymptotic freedom in 4 dimensions. In this case, as in ordinary
gauge theories \cite{G}, the only fixed point is $\lambda^*=0$.
Moreover, for $0<\lambda <1$ the IR trajectories will always go away
from $\lambda^*=0$, what is an indication of a first-order phase
transition in such theory.

However, taking into account that we are interested in the IR phase of
the
model, where the properties of the classical action are not important,
we may choose also $\lambda < 0$ \cite{EOpl94}. In this case, we have
the fixed point
\beq \lambda^*=-{ 60 \varepsilon (4 \pi)^2 \over 799}. \label{spl} \eeq
The general solution is
\beq
\lambda(t)={ \lambda e^{-\varepsilon t} \over
\ds 1-{799\lambda \over 60 \varepsilon (4 \pi)^2}(e^{\varepsilon t}-1)
},
\label{lambdat}
\eeq
which again shows the existence of the fixed point \req{spl} in the IR
($ t \to -\infty$).

\begin{table}

\begin{center}
\begin{tabular}{r|r|r|r|r|c|}
&$\lambda$&$\omega$&$\xi$&$f$&nature \\ \hline
1&-11.858&-0.298&-1.612&278.596& saddle point \\
2&-11.858&-0.025&0     &0      & " \\
3&-11.858&-0.023&0.201 &17.336 & " \\
4&-11.858&-0.023&0.203 &9.806  & IR stable fixed point \\
5&-11.858&-5.470&0    &0      & saddle point \\
6&-11.858&-5.470&0.026&33.108  & " \\ \hline
1&     0&arbitrary&   arbitrary&0&unstable point \\
2&     0&arbitrary&${1\over 6}$&${\varepsilon (4 \pi)^2 \over 3}$& " \\
\hline
\end{tabular}
\caption{ Numerical values for the fixed points of the
RG equations (3.6) (for $\varepsilon=1$). }
\end{center}
\end{table}

Table 1 is obtained by numerically solving equations \req{betasloxf}
with the r.h.s. set to zero.
There is only one IR stable fixed point.  Notice that the value
of $f$ corresponding to the stable fixed point is of the same order as
---but significantly smaller than--- in the absence of QG.
The remaining five fixed points are saddle
points (this is seen after careful numerical analysis of the matrix of
derivatives of the beta functions).
Thus, we observe that there exists an IR stable fixed point at which
a second order phase transition is taking place. Moreover, we get
a richer phase structure (four new fixed points) as compared with
the case of no QG.

Let us consider now scalar QED
\bea
{\cal L}_m&=&\ds
{1 \over 2}( \partial_{\mu}\varphi_1
-e \mu^{\varepsilon/2} A_{\mu} \varphi_2 )^2
+{1 \over 2}( \partial_{\mu}\varphi_2
-e \mu^{\varepsilon/2} A_{\mu} \varphi_1 )^2 \nn
&&\ds+{\xi \over 2} R \varphi^2
-{1 \over 4!}f \mu^{\varepsilon} \varphi^4-{1 \over 4}F_{\mu \nu}^2,
\eea
where $\varphi^2=\varphi_1^2+\varphi_2^2$.
One can write the RG equations for scalar QED with
$R^2$-gravity as:
\beq
\begin{array}{lllll}
\ds{d\lambda\over dt}&=&\beta_{\lambda}&=&\ds-\varepsilon\lambda
-{203 \lambda^2 \over 15 (4 \pi)^2}, \\
\ds{de^2\over dt}&=&\beta_{e^2}&=&\ds-\varepsilon e^2
+{2 e^4 \over 3 (4 \pi)^2}, \\
\ds{d\omega\over dt}&=&\beta_{\omega}&=&\ds
-{\lambda \over (4 \pi)^2}\left[ {10 \over 3}\omega^2
+\left( 5+{203 \over 15} \right) \omega + {5 \over 12}
+3 \left( \xi -{1 \over 6} \right)^2 \right], \\
\ds{d\xi\over dt}&=&\beta_{\xi}&=&\ds {1 \over (4 \pi)^2}
\left( \xi -{1 \over 6} \right)\left( {4 \over 3}f - 6e^2 \right)
+\Delta\beta_{\xi}, \\
\ds{df\over dt}&=&\beta_f&=&\ds-\varepsilon f
+{1 \over (4 \pi)^2}
\left( {10 \over 3}f^2 -12e^2f +36e^4 \right)
+\Delta\beta_f, \\
\end{array}
\label{betasleoxf}
\eeq
where $\Delta\beta_{\xi}$ and $\Delta\beta_f$ are defined in
\req{betasloxf}.
For $\lambda < 0$, $e^{* 2}={3 \over 2} \varepsilon (4 \pi)^2$
the numerical solution of the system for the fixed points
 shows no fixed point for $f$.
When $e^{* 2}=0, \lambda^*=0$, the situation is qualitatively the same
as in pure $f\varphi^4$-theory with $R^2$-gravity ---we get a few
unstable fixed points for $f$.

Now, we will discuss the $SU(2)$ gauge theory with scalars (in the
adjoint representation of the gauge group.
 Such a model is of importance because the IR behaviour of the
SM at finite temperature is described by the confining phase
of some $SU(2)$ gauge theory with scalars. Note that this theory is asymptotically free
(see \cite{1343,2706}).

The Lagrangian we are going to study is given by
\beq
{\cal L}_m=-{1 \over 4}G_{\mu \nu}^{a 2}
+{1 \over 2} (\nabla_{\mu}^{ab} \varphi_b )^2
-{f \mu^{\varepsilon} \over 4!} (\varphi_a^2)^2
+{\xi \over 2} R \varphi_a^2 ,
\label{LSU2}
\eeq
where
$\nabla_{\mu}^{ab}=\delta^{ab}\nabla_{\mu}
+\mu^{\varepsilon/2} g \epsilon^{acb} A_{\mu}^c, a=1,2,3$.
This theory, including also spinors, was studied in flat and curved
spacetime in the AF regime in refs. \cite{VT} and \cite{BOi},
respectively.

In the case of the $SU(2)$ theory, we get
\beq
\begin{array}{lllll}
\ds{d\lambda\over dt}&=&\beta_{\lambda}&=&\ds-\varepsilon\lambda
-{279 \lambda^2 \over 20 (4 \pi)^2}, \\
\ds{dg^2\over dt}&=&\beta_{g^2}&=&\ds-\varepsilon g^2
-{14 g^4 \over (4 \pi)^2}, \\
\ds{d\omega\over dt}&=&\beta_{\omega}&=&\ds
-{\lambda \over (4 \pi)^2}\left[ {10 \over 3}\omega^2
+\left( 5+{279 \over 20} \right) \omega + {5 \over 12}
+{9 \over 2} \left( \xi -{1 \over 6} \right)^2 \right], \\
\ds{d\xi\over dt}&=&\beta_{\xi}&=&\ds {1 \over (4 \pi)^2}
\left( \xi -{1 \over 6} \right)\left( {5 \over 3}f - 8g^2 \right)
+\Delta\beta_{\xi}, \\
\ds{df\over dt}&=&\beta_f&=&\ds-\varepsilon f
+{1 \over (4 \pi)^2}
\left( {11 \over 3}f^2 -24 g^2f +72g^4 \right)
+\Delta\beta_f . \\
\end{array}
\label{betaslgoxfSU2}
\eeq
The corresponding system for the fixed points
 does not have a stable solution. Only the unstable fixed
points $\lambda^*=0$ (when QG effects effectively disappear) or
 $g^{2 *}=0, \lambda^* < 0$, and some other associated
fixed points for $f$.

A few words about critical exponents are called for.
For $R^2$-gravity with matter we have few
mass-like terms; so, in principle, a calculation of the critical
exponent (more precisely, the critical-exponent matrix) is possible.
However, the explicit one-loop renormalization of massive terms in
 $R^2$-gravity with matter
has not yet been done.
Even if it were known, due to
the higher-derivative kinetic terms for the metric, and to the rather
nonstandard massive-like terms
for the gravitational field,
the interpretation of the critical-exponent matrix would still not be
clear. These
questions require further and careful study, and we hope to return to
them  elsewhere.

\section{RG-equations in higher-derivative QG with the $O(N)$
$\varphi^4$-theory}
In this section we will discuss the RG equations in the
$O(N)$ $\varphi^4$-model interacting with QG in
$D=4-\varepsilon$ dimensions. As we will see, by increasing the number
of scalars in the theory, we will be able to decrease the values of the
fixed points.

To start with, the Lagrangian of matter  is
\beq
{\cal L}_m=
{1 \over 2}g^{\mu \nu} \partial_{\mu}\varphi^i\partial_{\mu}\varphi^i
+{1 \over 2}\xi R \varphi^{i 2}
-{1 \over 4!}f \mu^{\varepsilon} (\varphi^{i 2})^2,
\label{LONphi4}
\eeq
where $i=1, \dots, N$.

For $R^2$-gravity interacting with the $O(N)$-theory, the RG
equations read
\beq
\begin{array}{lllll}
\ds{d\lambda\over dt}&=&\beta_{\lambda}&=&\ds-\varepsilon\lambda
-{\alpha^2 \lambda^2 \over (4 \pi)^2}, \
\alpha^2={133 \over 10}+{N \over 60}, \\
\ds{d\omega\over dt}&=&\beta_{\omega}&=&\ds
-{\lambda \over (4 \pi)^2}\left[ {10 \over 3}\omega^2
+\left( 5+\alpha^2 \right) \omega + {5 \over 12}
+{3N \over 2} \left( \xi -{1 \over 6} \right)^2 \right], \\
\ds{d\xi\over dt}&=&\beta_{\xi}&=&\ds {1 \over (4 \pi)^2}
{N+2 \over 3} f \left( \xi -{1 \over 6} \right)
+\Delta\beta_{\xi}, \\
\ds{df\over dt}&=&\beta_f&=&\ds-\varepsilon f
+{1 \over (4 \pi)^2}
{N+8 \over 3}f^2
+\Delta\beta_f , \\

\end{array}
\label{betasloxfON}
\eeq
where $\Delta\beta_{\xi}$ and $\Delta\beta_f$ are given by
\req{betasloxf}.
Numerical study of the fixed points of the system \req{betasloxfON}
for nonzero $\lambda$ and $\varepsilon=1$
yields the results in table 2.

\begin{table}
\begin{center}
\begin{tabular}{|r|r|r|r|r|c|}
\hline\hline
$N$&$\lambda$&$\omega$&$\xi$&$f$&nature \\ \hline \hline
10000& 0.88&-2.354&     0&       0& saddle point \\
     & 0.88&-1.141& 0.036&  0.0001& " \\
     & 0.88&-0.003& 0.163&  0.0084& " \\
     & 0.88&-0.003& 0.164&  0.0354& IR stable fixed point\\
     & 0.88&-0.031& 0.185&  0.0049&  saddle point \\
     & 0.88&-0.094& 0.200&  0.0041& " \\
     & 0.88&-53.135&     0&       0& " \\
     & 0.88&-54.896& 0.082&  0.0455& " \\ \hline
100000& -0.094&-2.485&     0&       0& saddle point \\
      & -0.094&-1.450&0.039 & 0.000001 &  " \\
      & -0.094&-0.002&0.162 & 0.000216 &  " \\
      & -0.094&-0.001&0.164 & 0.004044 & IR stable fixed point \\
      & -0.094&-0.003&0.184 & 0.000051 & saddle point \\
      & -0.094&-0.125&0.204 & 0.000041 &  "  \\
      & -0.094&-503.005&    0& 0       &  "  \\
      & -0.094&-504.868&0.083& 0.004717 &  "  \\  \hline
1000000&-0.0095&-2.4985&    0&        0 &  saddle point \\
       &-0.0095&-1.4540&0.039&  0.000000 &   "  \\
       &-0.0095&-0.0020&0.162&  0.000002 &   "  \\
       &-0.0095&-0.0001&0.165&  0.000439 & IR stable fixed point \\
       &-0.0095&-0.0266&0.183&  0.000001 &  saddle point \\
       &-0.0095&-0.1278&0.204&  0.000000 &   "  \\
       &-0.0095&-5002.9915&     0&      0&   " \\
       &-0.0095&-5004.8753&0.083&0.000474&  "  \\ \hline \hline
\end{tabular}
\caption{  Numerical values for the fixed points of the
RG equations (4.2).}
\end{center}
\end{table}

So, as we see in that table, for  $R^2$-gravity with
many scalars ($N \sim 10^4$ or larger) there exists an IR stable fixed
point where the values of all the coupling constants are less than one.
Hence, perturbation theory may be completely trusted around this IR
stable fixed point where a second-order phase transition at nonzero
temperature may be expected. We have in this case a few saddle
points too, where all the values of the couplings are also less than 1,
and
therefore a first-order phase transition may be expected around these
saddle points.

It is interesting to compare the $f^*$ corresponding
to the IR stable fixed point (when $N=1000000$) with the case of no QG.
In the latter,  we get
\beq
\left. {3 \varepsilon (4 \pi)^2 \over N+8}
\right\vert_{\varepsilon=1, N=1000000} = 0.000474
\eeq
(see the figure in the previous table, which gives 0.000439 for the same
quantity).
As we see, this value is changed (by about a $7 \%$) in the presence
of QG.  When $N$ decreases
this QG perturbation of the IR stable fixed point gets more and more
significant.
Thus, the QG perturbation of the IR-stable fixed point is quite a
visible effect even at very large $N$.
In the absence of
QG the critical exponent is quite well-known \cite{Wi}
\beq
{1 \over \nu}= 2 -{N+2 \over N+8}\varepsilon + O( \varepsilon^2 ) .
\eeq
Hence, at large $N$ we obtain the approximate value of the critical
exponent even in the presence of QG. It is clear from the above
discussion  that, for large $N$, QG corrections to this critical
exponent will be negligible.

\section{The RG-improved effective potential in three dimensions}
As an application of the above results, we will now study the
RG-improved
effective potential in three dimensions in the presence of QG.

We will consider the $SU(2)$ model. It is known that the IR behaviour
of the SM is determined by the confining phase of the $SU(2)$ theory.
Thus, these results may be useful for the study of the SM at nonzero
temperature.
First of all, we present the solution for $f(t)$ when QG is absent
(see also \cite{Wi} and \cite{BF}):
\beq
f(t)=g^2(t)\left\{ { \sqrt{2151} \over 11} \tan\left[
\arctan\left( {f\over g^2}-{15\over 11} \over {\sqrt{215} \over 11}\right)
-{\sqrt{239} \over 14}\log
\left( e^{\varepsilon t} {g^2(t) \over g^2} \right)
\right]
+{15 \over 19} \right\}
\label{ftnoQG}
\eeq
If QG corrections are small, this solution provides an approximate
description of the situation.
Going on to cases where QG corrections 
are significant, we will restrict
ourselves to a set-up in which $\lambda(t)$ is proportional to $g^2(t)$.
This is a very common situation, as there exist many GUT models with
$R^2$-gravity \cite{BS} (see \cite{BOS} for a review) which are
asymptotically free for all coupling constants, and where the QG
coupling $\lambda(t)$ is of the same order as the gauge coupling.

One can study (but only numerically) the running
coupling constant $f(t)$ for $R^2$-gravity. Of
course, the above discussion is somewhat speculative, as the way QG
corrections show up in the IR region is anything but clear. (Let us
recall, however, the proposed models of effective QG in the far infrared
\cite{AM} which may provide the framework for such large corrections).

We may also examine the situation where the QG corrections are large as
compared with scalar loops (gauge coupling corrections are usually so).
Then, one can write
\beq
{df \over dt}=-\varepsilon f
+ {1 \over (4 \pi)^2}\left( 72 g^4 + {5 \over 12}\lambda^2 \right), \ \
\ f(0)=f,
\eeq
which has  the following solution:
\beq
f(t)=e^{-\varepsilon t} \left\{
f
-{36 g^2 \over 7}
\left[ {1 \over \ds 1-{14 g^2 x \over \varepsilon} } -1 \right]
-{5 \lambda \over 167}
\left[ {1 \over \ds 1-{167 \lambda x \over 12} } -1 \right]
\right\} ,
\label{ftlQG}
\eeq
where
$\ds x= { e^{-\varepsilon t}-1 \over (4 \pi)^2 }$.
As one can see explicitly, even for $| \lambda | \simeq g^2$
($\lambda$ may be positive or negative in \req{ftlQG}) QG corrections to
$f(t)$ are very small. However, for $\lambda > g^2$,
QG corrections to \req{ftlQG} are still small
but now QG influences the dynamics in a different way. Indeed, for
$\lambda > g^2$, QG effects determine the Landau pole, which is reached
before the Landau pole associated to $g^2(t)$.

Now we may examine the RG-improved potential in the theory under
discussion. The standard way to write the RG equation for the effective
potential $V^{\varepsilon}(\varphi)$ is
\beq
\left( \mu{ \partial \over \partial \mu}
+\beta_{p_i}{ \partial \over \partial p_i}
-\gamma\varphi{ \partial \over \partial \varphi }
\right) V^{\varepsilon}(\varphi)=0,
\label{Ve}
\eeq
(for a recent, similar discussion without QG see
\cite{FJSE},\cite{BF},\cite{BKMN}, and in curved spacetime, see
\cite{EO}). where
$p_i$ are all our coupling constants,
the $\beta_{p_i}$'s denote
their corresponding $\beta$-functions, and $\gamma$ is the anomalous
scaling dimension for the scalar field $\varphi$.
Working on flat background one easily gets:
\beq
V^{\varepsilon}(\varphi)=
{ \mu^{\varepsilon} f(t) \over 4! } \varphi^4(t),
\label{VRGepsphi}
\eeq
where $\varphi^2=\varphi_a^2$, $f(t)$ (as well as $g^2(t)$,
$\lambda(t)$) are defined by eqs. \req{betaslgoxfSU2},
and $f(t)$ has
already been found (see above). The
effective field satisfies
\beq
{d\varphi(t) \over dt}= -\gamma(t) \varphi(t),
\label{dphit}
\eeq
 for $R^2$-gravity with $SU(2)$ model
\beq
\gamma={1 \over (4 \pi)^2} \left[  -6g^2
+{\lambda \over 4}\left(
{13 \over 3} -8\xi -3\xi^2 -{1 \over 6 \omega} -{2 \xi \over \omega}
+{3 \xi^2 \over 2 \omega}
\right) . \right]
\eeq
Then,
\beq
\varphi(t)= \varphi \exp\left[ \int_0^t dt  \gamma(t) \right] .
\label{phiti}
\eeq
Substituting \req{phiti} into \req{VRGepsphi}, taking $\varepsilon=1$
and using the explicit expressions for the couplings
one can find the leading-log behaviour of the effective potential
in three dimensions with account of QG corrections. Note that
$t={1 \over 2}\log{\varphi^2 \over \mu^2}$ and $\mu$ should be
identified with the temperature \cite{BF}.
Examples of the RG-improved potential thus obtained are shown
in Fig. 1,
for the $SU(2)$ model with and without QG.
The appearance of a
Landau pole near which perturbation theory breaks down is clearly seen
in these figures.

As one can realize, QG corrections to the $D=3$ RG-improved potential,
even in the $\lambda \simeq g^2$ case, are quite small. Here, we have shown
how QG corrections in the infrared behaviour of the SM can be in
principle taken into account ---through the study of the confining
phase of the $SU(2)$ theory. Despite the expectations that the
IR phase of QG could be important for particle phenomenology in the GUT
epoch, we have shown with this
simple example that QG corrections are actually not so essential
---while they
may, of course, still change the numerical values of the quantities
obtained in the absence of QG.

\section{Conclusions}

In this paper we have discussed renormalizable higher-derivative
QG with matter in $D=4-\varepsilon$ dimensions. The
RG equations have been written for
 $D=4-\varepsilon$ with different choices of
matter and the fixed points of these equations have been found. In the
example of $f\varphi^4$-theory with QG we show, by numerically solving
these equations, that the model possesses a rich phase structure, and
new fixed points for the scalar coupling constant ---with respect to
the case of no QG--- come into existence.
Unfortunately, no new IR stable fixed points appear as a result of QG
effects. The IR stable fixed point for the scalar coupling constant,
which was already present in the absence of QG, is
perturbed (about a
7 \%)  by QG effects. Near this IR stable fixed point, it is
predicted that the theory at nonzero temperature undergoes a
second-order phase transition.

For higher-derivative QG with scalar QED and with nonabelian gauge
fields, we obtain only unstable fixed points, as solutions of the RG
equations. We have also considered the $O(N) \varphi^4$-theory, where
gravitational corrections to fixed points get smaller when we increase
the number of scalars ---as it should be. In this case, the IR stable
fixed point, perturbed by QG,  exists as well.

We have studied the $SU(2)$ gauge model with scalars interacting with
higher-derivative QG and calculated the RG-improved effective potential
at nonzero temperature in that theory (which approximately describes the
confining phase of the standard model). Numerical estimates show that QG
corrrections can give some nonessential contributions to this potential,
which can actually become sizeable near the Landau pole (at least for
some values of the gravitational coupling $\lambda$).

Recently, $R^2$-gravity without matter was investigated in
$2+\varepsilon$
dimensions \cite{NTT}.
It would be of interest to try and combine the
$4-\varepsilon$ and $2+\varepsilon$ approaches in $R^2$-gravity, which
we have attempted to formulate here, in order to obtain somehow
coinciding results at $\varepsilon=1$. Such a synthesis might certainly
enrich  both approaches very much.

In summary, for the system consisting in renormalizable $R^2$-gravity
with a scalar-gauge theory near four dimensions, we have shown that the
$\epsilon$ expansion technique can be applied to study the critical
behavior of the theory and its temperature phase transitions. This model
provides an explicitly workable example of how four-dimensional QG could
actually influence some well established results \cite{G} on temperature
phase transitions and IR fixed points.

\vskip1cm
\ni{\large\bf Acknowledgements}

 S.D. Odintsov thanks
I. Antoniadis and R. Tarrach for  discussions. This work has been
supported by DGICYT (Spain), project  PB93-0035, by CIRIT
(Generalitat de Catalunya) and by RFFR, project 94-02-03234.

\newpage
\ni{\large\bf Figure Captions}
\vskip1cm

\ni{\bf Fig. 1} a) RG-improved effective potentials for
the $SU(2)$ model
with $\omega (0) =0$, $\xi (0) = 1/6$,  $g^2(0)=0.41$, $f(0)=0.5$ and $\mu=1$.
The three lines drawn correspond to $\lambda(0)=$0, 0.4 and 0.8, as
indicated by the labels. b) Same picture for the theory with
$\omega(0)=1, \xi(0)=0.5$,
and taking $\lambda(0)=10^{-3}$, 0.4, 0.8.

\end{document}